\documentclass{epl}

\title{Percolation in real Wildfires}

\author{Guido Caldarelli\inst{1}, Raffaella Frondoni\inst{2,}\inst{3}
Andrea Gabrielli\inst{1}, Marco Montuori\inst{1}, Rebecca Retzlaff\inst{4} 
\and Carlo Ricotta\inst{3}}
\institute{
  \inst{1} Sezione INFM di Roma1 and Dipartimento di Fisica,
Universit\`a "La Sapienza", P.le A.Moro 2 00185 Roma, Italy.\\
  \inst{2} Department of Geography, University of Cambridge,
Downing Place, Cambridge, CB2 3EN, UK \\
  \inst{3} Dipartimento di Biologia Vegetale, Universit\`a di Roma
"La Sapienza", P.le A. Moro 2 00185 Roma, Italy\\
  \inst{4} Remote Sensing Department,
University of Trier, Behringstrasse 54286 Trier Germany
}
{Structures and organization in complex systems} 

\begin{document}

\maketitle

\begin{abstract}
This paper focuses on the statistical properties of wild-land fires and, 
in particular, 
investigates if spread dynamics relates to simple invasion model.
The fractal dimension and lacunarity of three fire scars classified from 
satellite imagery are analysed. 
Results indicate that the burned clusters behave 
similarly to percolation clusters on boundaries and look more dense in 
their core.  We show that Dynamical Percolation reproduces this behaviour and
can help to describe the fire evolution. By mapping fire dynamics
onto the percolation models the strategies for fire control might be 
improved. 
\end{abstract}
\vspace{1cm}

In recent times the introduction of satellite imaging facilitated the 
coarse-scale analysis of wildfires\cite{ricotta}.
Being inspired by the self-similar aspect of the fire scars, we want here to 
provide an explanation for this lack of characteristic length scale.  
We want to link wildfires spreading with the evolution of diffusive systems whose
scale invariance has been widely analyzed\cite{baffo}.
In particular we focus here on the simplest example of 
fractal growth model, the model of percolation\cite{perc}. 
Percolation has been extensively studied, and  
it proved to be extremely successful in explaining some of the statistical 
properties of
several propagation phenomena ranging from the polymer gelation\cite{flory}
to superconductors\cite{norm}.

In this paper we present some evidence that percolation models could be
fruitfully applied to describe properties of wildfires\cite{lit}.
In particular, the dynamical version of percolation, known as Dynamical
Percolation (DyP), may provide effective insights into the fire control.
We based our analysis on the comparison between several statistical properties
of wildfires with those of the percolation clusters.
We report here the measures of the 
fractal dimension of areas, {\em accessible} perimeters and hulls 
(defined as the set of the
most external sites of the cluster)\cite{perc}, 
along with the lacunarity (i.e. the void distribution inside the cluster).
As a result, we can conclude that, within the error bars, statistical
properties of wildfires can be accurately described by 
a self-organized version of DyP \cite{dyp}. 

\begin{figure}
\onefigure{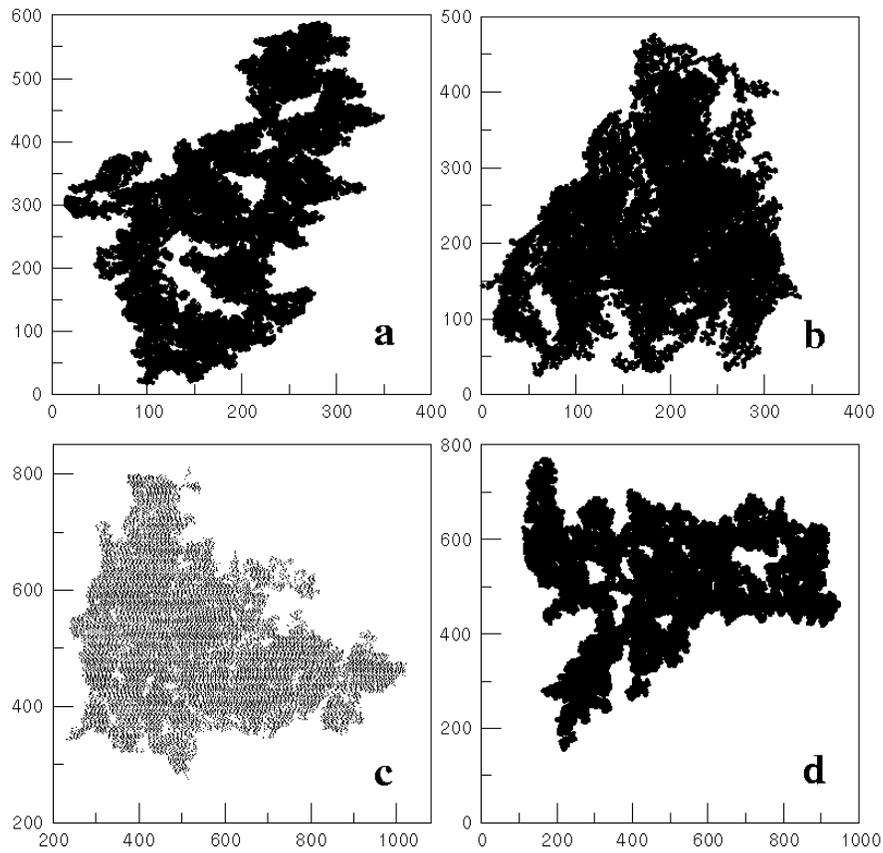}
\caption{Binary map of the burnt areas 
{\bf (a)} for the valley of Biferno,
{\bf (b)} for the Penteli wildfire,
{\bf (c)} for the Cuenca wildfire.
In {\bf (d)} we plot a cluster of Self-Organised Dynamical Percolation
whose dimensions are comparable with case {\bf (c)}.
Each pixel corresponds to an area of $900 m^2$.}
\label{fig1}
\end{figure}

The data set shown  here, consists of Landsat TM satellite imagery
($30$m $\times 30$m ground resolution) of wildfires, acquired respectively:
over the Biferno valley (Italy) in August 1988; over the Serran\'{\i}a Baja
de Cuenca (Spain) in July 1994; and over the mount Penteli (Greece) in July 1995.  
In all the cases the image was acquired a few days after fire. 
The burnt surfaces were respectively $58$, $60$, $156$ square Kilometers.  
Bands TM3 (red), TM4 (near infrared) and TM5 (mid infrared) of the post-fire 
subscene are classified using an unsupervised algorithm and 8 
{\em classes}\cite{desc2}.
This means that in the above three bands any pixel of the image is characterised by a 
value related to the luminosity of that area. By clusterising in {\em classes} those values
one can describe different type of soil, and in particular the absence or presence
of vegetation. 
In particular the maps of post-fire areas have been transformed into binary maps 
where black corresponds to burned areas. These maps are shown in Fig.\ref{fig1}.  

In order to quantify the possible scale
invariance, we measure the following properties:
{\bf(1)} the fractal dimension of the burned area;
{\bf(2)} the fractal dimension of the {\em accessible} perimeter;
{\bf(3)} the fractal dimension of the hull (defined as the
set of burned sites on the boundary of the system);
{\bf(4)} the variance of the relative point density fluctuations
(i.e. a measure of the lacunarity of the system).
We compare these values with the corresponding ones of 
self-organized Dynamical Percolation that we believe to represent the phenomenon. 

\begin{figure}
\onefigure{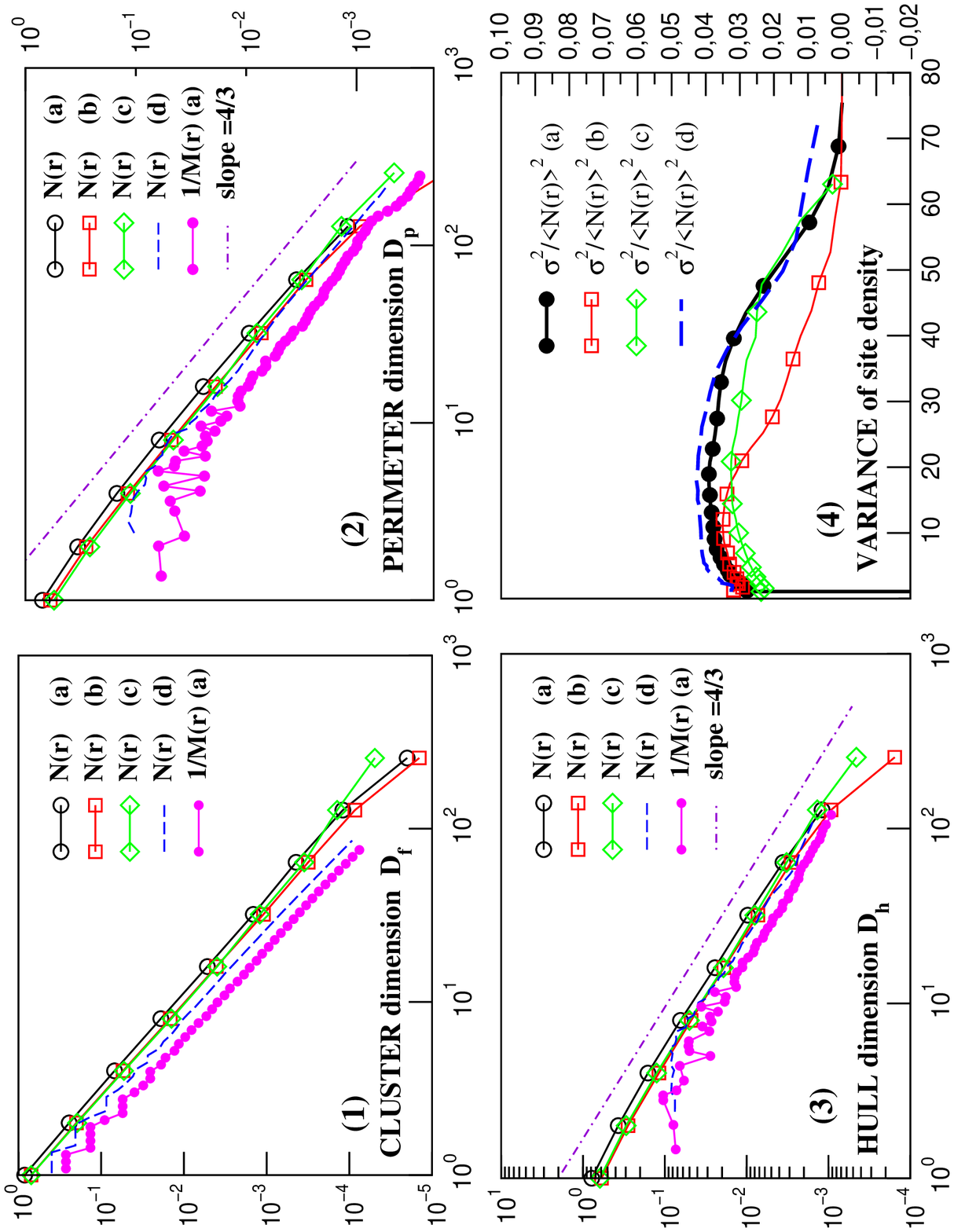}
\caption{Scatter plot of {\bf(1)} the fractal dimension of the aggregate,
{\bf(2)} the fractal dimension of the perimeter of the structure,
{\bf(3)} the fractal dimension of the hull of the structure,
{\bf(4)} the variance in the relative site density fluctuations. 
In all the plots we distinguish between cases {\bf a,..,d}. 
For the fractal dimensions we report for case {\bf a} both the measures 
$N(r)$ and $1/M(r)$.}
\label{fig2}
\end{figure}

To measure fractal dimensions we apply two different methods:
the average mass-length relation and the box counting method.
Box counting method is performed overlapping a grid of 
size $r$ over the data set and counting the number $N(r)$ of boxes occupied 
by the cluster at the scale $r$.
For fractal objects 
$N(r)\propto r^{-D_f}$ (for $r\rightarrow 0$) 
where $D_f$ is the fractal dimension.
The average mass-length relation measures
the relationship 
between the average number $M(r)$ of points of the date set at distance $r$ 
around any other point of the data set itself.
This can be achieved, by measuring $M(r)$ 
in a circle of radius $r$ centered around a point of the system. 
For scale-free objects $M(r)\propto r^{D_f}$ (for $r\rightarrow \infty$).
To avoid any bias in the result, circles should be fully included in the 
cluster.
The two methods gave equal results within the error bars.
In general this means that $D_f$ is a well defined property of the system.
These results for the box counting are shown in Fig.\ref{fig2} and 
summarized in table \ref{tab1}.
This table also reports the exact values that characterize 
critical percolation clusters.
All wildfire measures but the hull are in very good agreement with percolation 
data. 

\begin{figure}
\onefigure{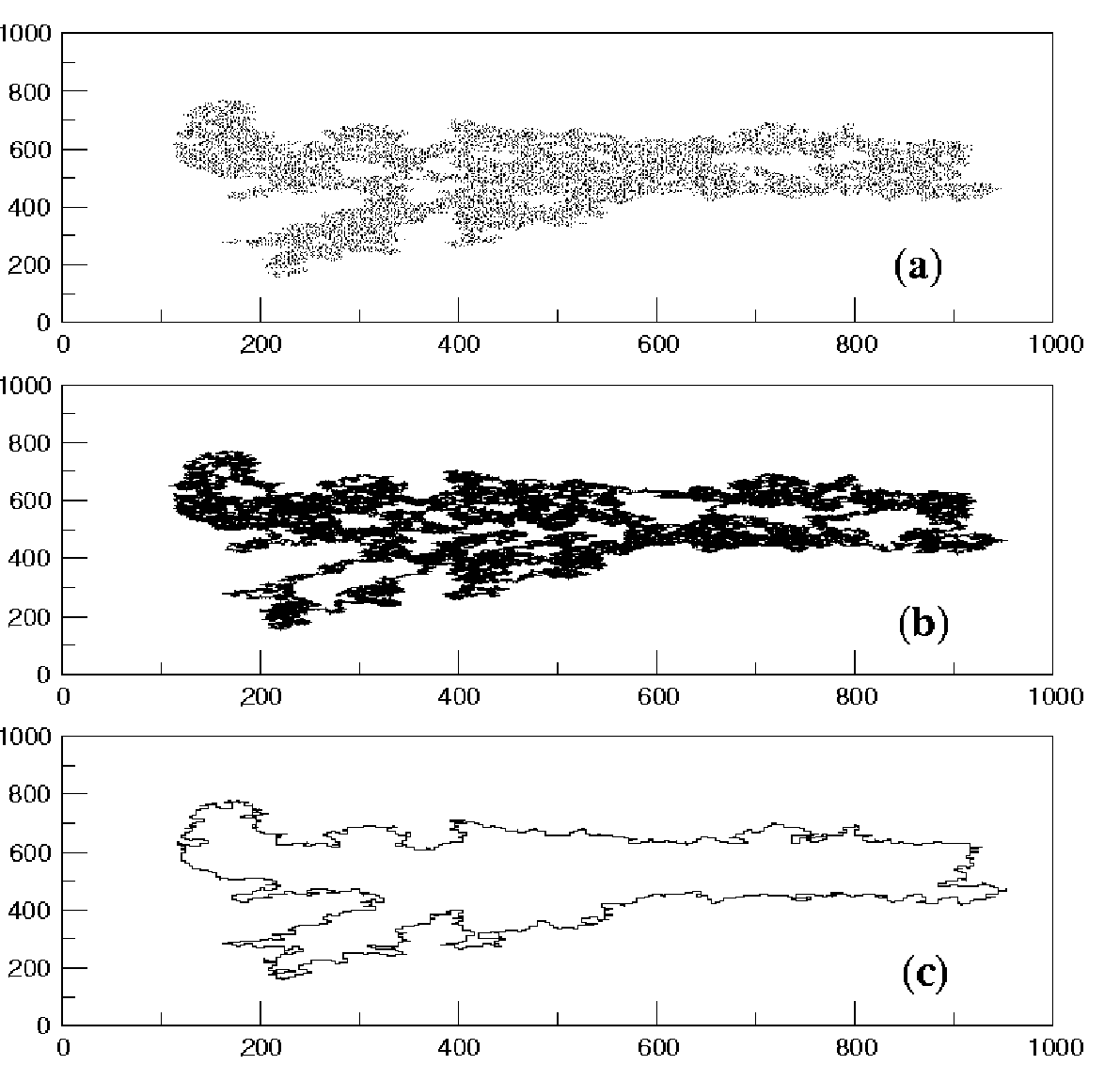}
\caption{Numerical Simulation of an Dyp cluster.
In (a) we show the cluster, in (b) the hull of this cluster. in (c) we show
the hull of the coarse-grained picture of cluster in plot (a).
The coarse-grained version is obtained by using cells with linear size twice
of the original one. After only two steps the statistical 
properties of the hull become similar to those of the perimeter.}
\label{fig3}
\end{figure}

We believe that this peculiar behaviour for the hull may depend 
on the coarse resolution of the remotely sensed burnt area ($30$ m), resulting
in a kind of Grossman-Aharony effect \cite{grossman} which reduces the hull
of the critical percolation cluster to the accessible perimeter. 
Consequently, the hull
fractal dimension (equal to $7/4$) is reduced to the fractal dimension 
of the accessible perimeter (equal to $4/3$).
This effect can be induced by different redefinition of the connectivity 
criterium on the hull sites.
In this case in particular we refer to the studies of Ref.\cite{Kolb,Kolb2} where 
a smaller than expected hull exponent is related to the low resolution 
of image with respect to the characteristic distance of the percolation 
process (see also Ref.\cite{gbs}). 
To test this assumption, a critical percolation cluster computer simulation 
with $3\times 10^5$ sites was undertaken. 
Results show that, whereas most statistical
properties do not change, the hull tends to
behave as the accessible perimeter if a coarse graining procedure is applied 
in such a way to reduce the resolution between first and second neighbors 
(see Fig.\ref{fig3}).

The last measure we perform is the computation of the
variance $\sigma(r)$ of the normalized point density fluctuations
of the burnt sites, i.e. $\sigma(r)= \sqrt{<M(r)^2>/<M(r)>^2-1}$.
Generally $\sigma(r)$ is a function of the radius $r$; for a
``simple'' fractal, the variance $\sigma(r)$
is a constant $\sigma_{intrinsic}$ .
At large scale $r$, for values near to the spatial
extension of the data set, the measure of statistical quantities
($M(r), \sigma(r)$ etc.) is affected by finite size effects.
Such effects produce
a decrease of the $\sigma(r)$
for increasing values of $r$.
As we can see in Fig.\ref{fig2}{\bf d}, the measures of the
$\sigma(r)$ versus $r$
both for the fire data and the computer simulated percolation cluster are in
good agreement.
Moreover, they fit the values reported in Ref.\cite{BCF}, where
the same quantity is estimated for an ordinary percolation cluster.
The value $\sigma^{2}_{intrinsic}$ can be considered as a measure of the
morphology of a fractal data set.
The larger is $\sigma^{2}_{intrinsic}$, the larger is the probability
that the fractal set has large voids.
This is evident from Fig.\ref{fig2}, where the variance has lower
values for the the wildfires with smaller voids (set {\bf b} and {\bf c}).

As a last remark on the data interpretation, 
we checked that the fuel load distribution before fires was rather uniform
in the analyzed areas. Therefore, we can exclude that fractal properties of 
fire depend on pre-fire vegetation distribution. 
Nevertheless additional work is underway to quantify the effects of pre-fire 
fuel load distribution on fire behaviour. 

 From the above data analysis it seems that percolating clusters could
describe reasonably well the process of fire spreading. 
Unfortunately in the original formulation, percolation is a static model 
where one considers sites on a lattice
that can be selected with a certain probability $p$.
If $p=1$ all points are selected and there are plenty of spanning paths
in the system.
When $p=0$ no point is selected and there is no way to form a spanning path.
By increasing $p$ step by step from zero,
small clusters of connected areas are generated,
until for a particular value of probability $p=p_c$, called
{\em percolation threshold}, a part of the small clusters
coalesce and form a spanning cluster.

Even if most of the properties measured in real fires are reproduced by percolation 
we need a model whose dynamics could mimic in a reasonable way that of the wildfires.
We then propose here to use the Self-Organised version of Dynamical Percolation.  
Dynamical percolation\cite{Jan} was introduced to study the 
propagation of epidemics in a population and its definition is the following:
each site of a square lattice can be in one of
three possible states:
(i) ignited sites, (ii) green sites susceptible to be ignite in the future,
and (iii) immune sites (i.e. burned sites non susceptible to be ignite again).
At time $t=0$ a localized seed of ignited sites is
located at the center of an otherwise empty (green)
lattice. The dynamics proceeds in discrete
steps either by parallel or by sequential updating
as follows: 
at each time-step every ignited site can
ignite a (green) randomly chosen
neighbor with probability $p$ or, alternatively, burn completely and
become immune to re-ignition with complementary
probability $1-p$. 
Any system state with no burning site is an 
{\it absorbing configuration}, i.e.,
a configuration in which the system is 
trapped and from which it cannot escape \cite{munoz, Granada}.
It is clear that
depending on the value of $p$ the fire generated
by the initial ignited seed will either spread
in the lattice (for large values of $p$) or die out
(for small values of $p$).
The two previous phases, are divided at the percolation threshold $p_c$, 
where the fire propagates marginally,
leaving behind a fractal cluster of immunized sites. 
Interestingly, it can be shown using field theoretical tools
that this is a critical percolation cluster\cite{Jan}.
In this way we have a dynamical model which, at criticality,
reproduces the (static) properties of standard percolation.
Clearly this model presents extensive fractal properties only if $p=p_c$.
The tuning of this parameter exactly to $p_c$ is however quite unlikely. 

For that reason we present here then a {\em self-organized} version of this 
model assuming a time-dependent form for the ignition probability
$p(t)$ decreasing from an initial value $p_0>p_c$ with time constant 
$\tau$ (e.g. $p_0\exp (-t/\tau)$ or $p_0/[1+(t/\tau)^n]$).
In the optics of fires, $p_0$ represent the initial ``force'' of the fire, and
$\tau$ is its characteristic duration. 
This observation came from the experience in fire control. Even without human 
activity fires eventually stops. 
It is then fair to introduce a fire extinction 
probability increasing with time. 
Fire will then invade new regions and will be able to continue 
until the percolation
probability is larger or equal to the critical value $p_c$.
This peculiar process is also able to reproduce in a qualitative way the results
of the fire clusters. Indeed the fire will grow almost in a compact way at the beginning, 
leaving a fractal boundary at the end of the activity.

Let us suppose to start the dynamics at $p_0>p_c$.
At the beginning the dynamics is the same of DyP with constant $p>p_c$, i.e.
the ignited region is quite compact, leaving only small holes of vegetation. 
However, with the time passing $p(t)$ reduces and then the diameter of 
islands in the burning cluster increases.
Finally, after a certain time
proportional to $\tau$ one has $p(t)<p_c$, then the dynamics arrests 
{\em spontaneously} in some time-steps.
In particular one can see that if $\tau\gg 1$, at the arrest time $t_f$,
$p(t_f)\simeq p_c$.
Therefore, the geometrical features of the final burnt cluster become 
more and more irregular (fractal) going towards the hull, representing an 
effective spatial (radial) probability of ignition.
One can show that the final hull and the accessible perimeter
have the same fractal dimensions for ordinary percolation.
However, this fractality is extended only up to a characteristic scale
$\xi\sim\tau^{\alpha_{\xi}}$ with $\alpha_{\xi}=1/D_h$.
$\xi$ gives also the characteristic scale of the voids nearby the hull, which
are the largest in the cluster.  
Moreover, $p_c-p(t_f)\sim \tau^{-\alpha_p}$ where $\alpha_p=(D_h-1)/D_h$.
In few words, the hull presents the main features find in another static
percolation model known as Gradient Percolation \cite{GP}.
We believe that these properties of the self-organized DyP
explain why in the largest analyzed wildfires (i.e 
starting with a larger $p_0$ or a larger $\tau$) we have an effective 
increase of the global fractal dimension $D_f$ towards $2$
and the appearance of large voids only nearby the hull.
Instead for the smallest ones one can think that $p_0$ is too near $p_c$
(with respect to the value of $\tau$) to realize the spatial 
gradient of ignition probability. This would result in fires cluster 
as large as $\xi$, and then with the same features of ordinary clusters
of percolation near criticality.

The importance of such a result lies in the particular growth dynamics
shown by DyP.
As pointed out in Ref.\cite{dyp,munoz} 
DyP grows mainly by selecting sites newly added 
to the system. If this applies also to the external boundaries of a
wildfire, one could in principle think to focus the activities of fire 
control where the fire invasion is faster.
Here it is also worth discussing the features of the so called ``forest fire
model'' (see \cite{Jen} for a complete review on the model).
In this model 
at successive time steps trees (sites) are removed trough simple rules of ignition
from nearest neighbours or by burning through external lighting.
At the same time new trees grow on the empty sites left by the fire.
This model not only presents the unrealistic assumptions of fast re-growing
trees in the system, but also produces almost compact clusters of wildfires
that fail to reproduce the statistical properties we observe in the data.
We believe that despite the name, this model fails to reproduce the behaviour 
of real wildfires representing instead a nice statistical model to show the
properties of Self-Organised systems.

In conclusion, we present here some of the statistical properties of wildland fires.
Results indicate that the cluster formed by fire shows at least on 
boundaries well defined fractal 
properties strictly related to percolation at criticality. 
In particular we believe that the most suitable 
model to describe fire dynamics is a self-organized version of 
Dynamical Percolation.
Unfortunately, due to the coarse spatial and temporal resolution of 
available satellites, it is very difficult to check the dynamical 
properties of the forest fires. 
Nevertheless, the very good agreement between DyP and real data
and the similar evolution of growth suggests that DyP could indeed 
represent a suitable model in most cases.
The assignment of random probability values to the links between 
sites, which is performed 
in the model construction, can effectively model the broad-scale 
cumulative effect of interacting features (terrain, vegetation, etc.). 
Since the dynamical properties of DyP have been extensively studied
recognition of DyP dynamics for fire spread has important consequences for 
fire control. This should focus
on the latest zones attacked by fire, since the zones left behind have a 
small probability to keep the fire alive. 
It is indeed a well known result that DyP grows through the most recent
areas entered in the system.

This work has been supported by the EU Contract No.FMRXCT980183.
This work was also partially supported by the LUCIFER (Land Use Change Interactions with 
Mediterranean Landscapes) project of the European Union (ENV-CT96-0320).

\begin{table}
\begin{centering}
\protect \caption{Fractal dimensions, for the data and for the DyP 
model. Exact values for DyP are computed on hierarchical lattices.}
\label{tab1}
\begin{tabular}{|c|c|c|c|c|}
\hline
$ $        & {\bf a} Biferno &{\bf b} Penteli&{\bf c} Cuenca &{\bf d} DyP \\ \hline
$D_f$      & $1.90(5)$ & $1.93(5)$ & $1.95(5)$  & $91/48$    \\
$D_h$      & $1.30(5)$ & $1.32(5)$ & $1.31(4)$  & $7/4$      \\
$D_p$      & $1.30(5)$ & $1.33(4)$ & $1.34(5)$  & $4/3$      \\
${\cal L}(1)$ & $0.037(3)$& $0.036(3)$& $0.034(3)$ & $0.040(2)$ \\
\hline
\end{tabular}
\end{centering}
\end{table}


\begin{thebibliography}{99}
\bibitem {desc2}
To evaluate robustness of results, classification has been repeated with different 
algorithms and number of classes (ranging from $6$ to $10$). 
Mapping of fire scar was mostly unchanged, proving high robustness of 
the spectral signature of fire.
Images of pre- and post-fire areas were then processed 
to conform to each other.
Transformation was based on measurements of ten ground control points,
that could be located unambiguously on both images. 
The maximum RMS error was 0.31 (around 10 meters); the mean RMSE was 0.17.
\end{thebibliography}
\end{document}